\documentclass[reprint,
 amsmath,amssymb,
 aps,]{revtex4-2}

\usepackage{graphicx}
\usepackage{dcolumn}
\usepackage{bm}
\usepackage{amsfonts}
\usepackage{amsmath}
\usepackage{amssymb}
\usepackage{color}
\usepackage{float}

\begin{document}


\title{Contagion in simplicial complexes}

\author{Z. Li}
\affiliation{School of Automation, Northwestern Polytechnical University, Xi'an 710072, China}

\author{Z. Deng}
\affiliation{School of Automation, Northwestern Polytechnical University, Xi'an 710072, China}

\author{Z. Han}
\affiliation{School of Cybersecurity, Northwestern Polytechnical University, Xi'an 710072, China}

\author{K. Alfaro-Bittner}
\affiliation{Departamento de F\'{i}sica, Universidad T\'{e}cnica Federico Santa Mar\'{i}a, Av. Espa\~{n}a 1680, Casilla 110V, Valpara\'{i}so, Chile}

\author{B. Barzel$^{*,}$}
\affiliation{Department of Mathematics, Bar-Ilan University, Ramat-Gan, 5290002, Israel}
\affiliation{Gonda Multidisciplinary Brain Research Center, Bar-Ilan University, Ramat-Gan, 5290002, Israel}

\author{S. Boccaletti$^{*,}$}
\affiliation{Universidad Rey Juan Carlos, Calle Tulip\'{a}n s/n, 28933 M\'{o}stoles, Madrid, Spain}
\affiliation{CNR - Institute of Complex Systems, Via Madonna del Piano 10, I-50019 Sesto Fiorentino, Italy}
\affiliation{Moscow Institute of Physics and Technology (National Research University), 9 Institutskiy per., Dolgoprudny, Moscow Region, 141701, Russian Federation}

\date{\today}

\begin{abstract}
The propagation of information in social, biological and technological systems represents a crucial component in their dynamic behavior. When limited to pairwise interactions, a rather firm grip is available on the relevant parameters and critical transitions of these spreading processes, most notably the pandemic transition, which indicates the conditions for the spread to cover a large fraction of the network. The challenge is that, in many relevant applications, the spread is driven by higher order relationships, in which several components undergo a group interaction. To address this, we analyze the spreading dynamics in a simplicial complex environment, designed to capture the coexistence of interactions of different orders. We find that, while pairwise interactions play a key role in the initial stages of the spread, once it gains coverage, higher order simplices take over and drive the contagion dynamics. The result is a distinctive spreading phase diagram, exhibiting a discontinuous pandemic transition, and hence offering a qualitative departure from the traditional network spreading dynamics.

\end{abstract}


\maketitle
$^{*}$ These Authors equally contributed to the Manuscript.
\\Corresponding author: k.alfaro.bittner@gmail.com


Networks represent a powerful tool to track complex spreading processes, from the spread of epidemics or ideas in social networks \cite{barrat2008dynamical,pastor2015epidemic,pastor2001epidemic,boguna2002epidemic}, to the propagation of failures in infrastructure systems \cite{boccaletti2006complex,watts2011simple,simonsen2008transient}. The network captures the underlying geometry of the spread, drawing links between all directly interacting components, and the dynamic interactions between these components govern the time-scales and spatial patterns of the spread, as it propagates along these links \cite{newman2018networks,newman2003structure,wu2015emergent,cowan2004network,Hens2019network,Harush2017Patterns}. This pairwise modeling framework, however, cannot account for higher order interactions \cite{centola2007complex,benson2016higher,grilli2017higher,guilbeault2018complex}, which often play a crucial role in social \cite{centola2010spread,ugander2012structural}, biological \cite{chan2013topology,nanda2014simplicial} or technological contagion processes \cite{de2007coverage,barbarossa2016introduction,pokorny2016topological}. For example, in social systems ideas are propagated both by direct interactions between individuals, but also, at the same time, by more complex social structures, governed by \textit{group} interaction, \textit{e.g}., when an individual is encouraged to adopt an idea following his or her social \textit{circle}. Similarly, in technological networks, the state of a specific component is often impacted by the cumulative failures of its surrounding units. Hence a single failed neighbor has little impact, but a combination of several failures induces additional stress, potentially leading to the component's subsequent failure.

\begin{figure}[t]
\centering
\includegraphics[width=8.5cm]{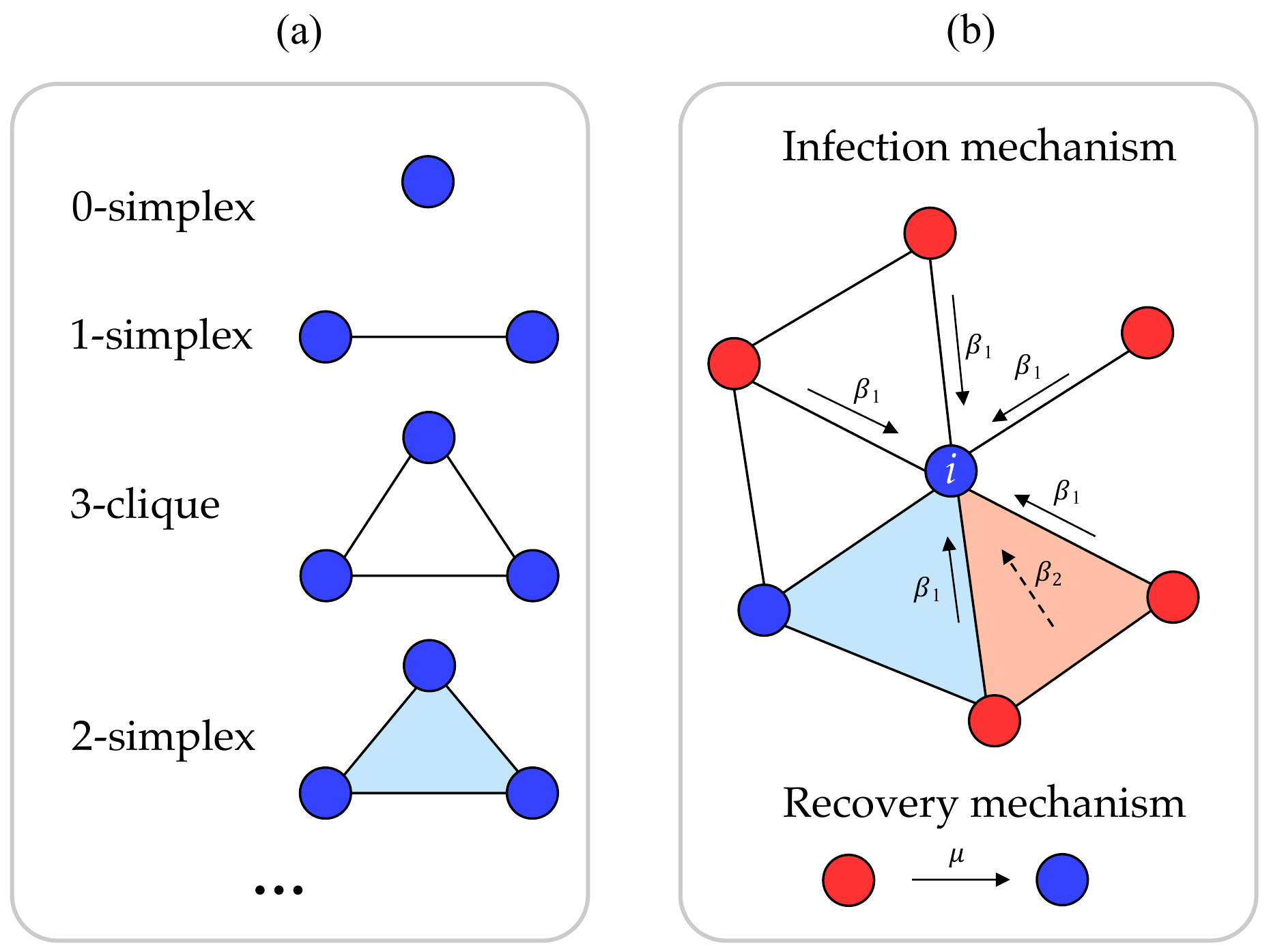}
\caption{(a) The underlying units of a simplicial complex. Simplices are different from traditional cliques and are distinguished with colored structures. (b) The mechanism of simplicial SIS spreading model. Blue and red nodes denote nodes in the susceptible and infected state, respectively. Infections occur in two ways. In non-simplex structures, node $i$ can be infected by its infected neighbors only through edges at rate $\beta_1$, which is the same as the traditional pairwise interaction. In simplex structures, taking a 2-simplex as an example, in addition to being able to be infected by its infected neighbors through edges at rate $\beta _1$, if two other neighbors are both infected, node $i$ can be also infected by a higher-order way at rate $\beta_2$, which is called higher-order interaction (or group interaction). The recovery mechanism is the same as the traditional case that infected nodes will recover at rate $\mu$.}
\label{fig1}
\end{figure}

To model such higher order interactions we must advance beyond \textit{networks}, and consider \textit{simplices} of three or more interacting components \cite{battiston2020networks,boguna2021network,wu2015emergent,mulder2018network}. For example, a $p$-simplex describes a simultaneous interaction between $p + 1$ nodes, $n_0,n_1,\dots,n_p$, which undergo a \textit{group} interaction, in which, \textit{e.g}., node $n_0$ is affected by the collective state of all nodes $n_1,\dots,n_p$. Hence, a $p$-simplex is \textit{not} just a clique in which all potential pairwise interactions are active, but a higher order, $p$-wise form of interaction, as one can see in the sketch of Fig.~\ref{fig1}(a).

To understand the patterns spread in a simplicial environment we implement the most fundamental contagion process - the susceptible-infected-susceptible (SIS) model \cite{anderson1992infectious,boguna2013nature,pastor2015epidemic,boccaletti2006complex,ferreira2012epidemic}, in which nodes can be either infected directly by their neighbors at a rate $\lambda_1$, a pairwise process, but also by their $p$-simplices at rate $\lambda_2$, capturing an order $p$ interaction \cite{iacopini2019simplicial,matamalas2020abrupt}. In a social system, for example, this describes the simultaneous effect of individual \textit{one-on-one} interactions, together with the influence of peer-pressure, exerted by each person's social circle.

We find that even under this simple contagion process, the presence of higher-order interactions fundamentally changes the patterns of spread. Specifically, we observe and analyze three distinctive fingerprints of simplicial contagion:\ (i) The spread exhibits two time-scales, the first, at the early stages, mediated by pairwise interactions, and hence dominated by $\lambda_1$; the second, ignited once the prevalence is sufficient to activate the higher order simplices, therefore governed by $\lambda_2$. (ii) The system features a discontinuous transition, characterized by a hysteresis phenomenon, as opposed to the classic, second order, pandemic transition of the traditional SIS model. (iii) We find that the pairwise interactions are crucial to \textit{seed} the contagion, but once the spread has covered sufficient ground, it can be sustained by the $p$-order interactions alone. In a social context, the meaning is that individual interactions are crucial to ignite the spread of an idea, but once it gains acceptance, peer-pressure alone can sustain its prevalence.

\begin{figure*}
\centering
\includegraphics[width=\linewidth, scale=1.00]{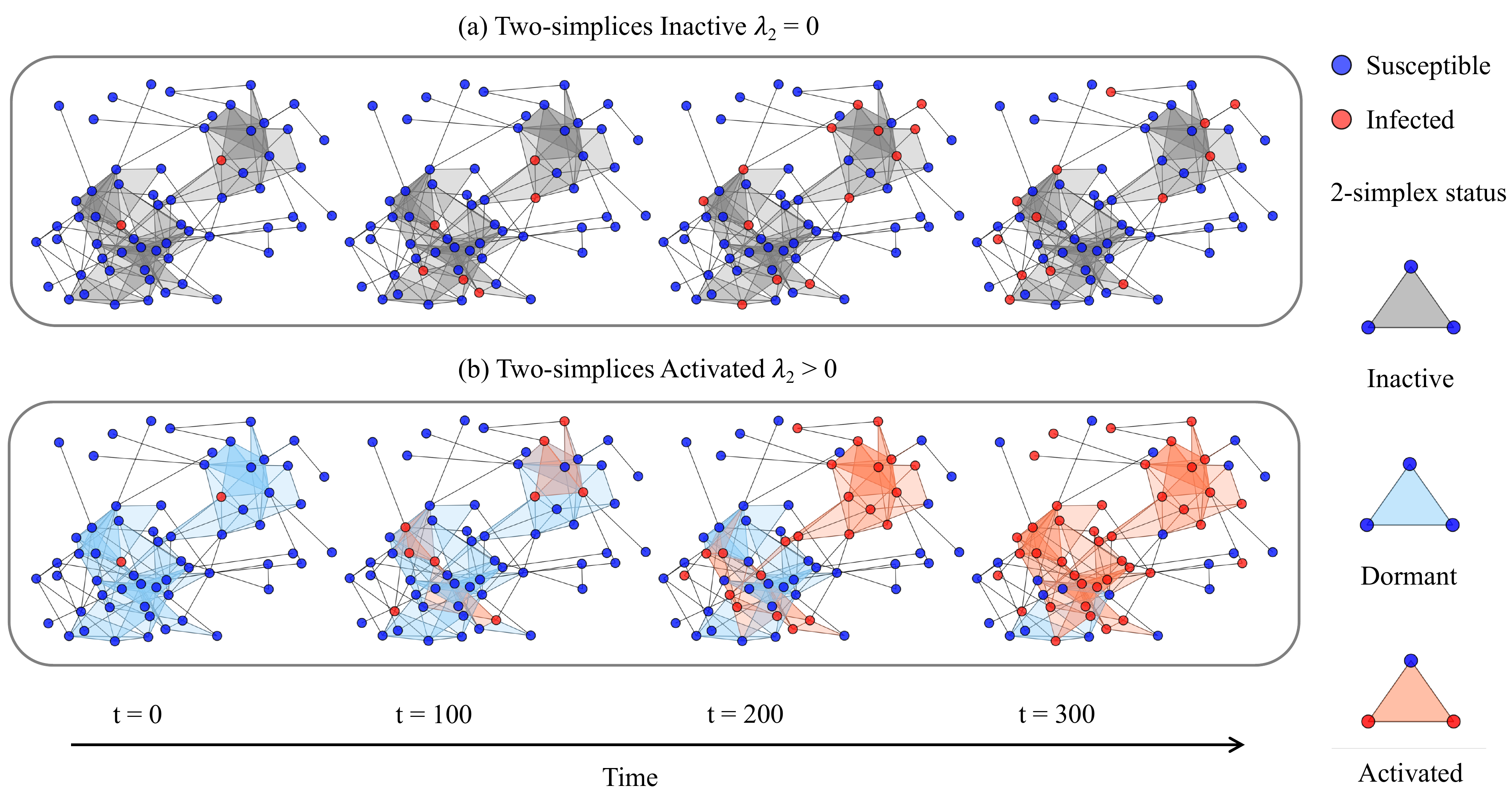}
\caption{\label{fig2} Evolutionary process of simplicial spreading on dolphins network. (a) The traditional SIS spreading process, where blue and red nodes are susceptible and infected respectively. Triangles denote the higher-order connections, and the grey-colored here just serve to show the structure, which actually has nothing to do with the SIS spreading process as the traditional cliques. (b) The simplicial spreading process, where blue- and red-colored triangles denote dormant and active 2-simplices, respectively. Higher-order interactions can only happen in an activated simplex, and a dormant $p$-simplex becomes activated if and only if $p$ nodes have been infected in the simplex. Dolphins network contains 62 nodes and 159 edges. In our simulations, we take a certain portion of triangles as 2-simplices which are selected randomly. $\lambda_2$ denotes the normalized infectious rate for higher-order interactions in triangles, which can be defined as $\lambda_2=\beta_2\langle k\rangle/\mu$, where $\langle k\rangle$ denotes the average node degree.}
\end{figure*}

Let us then describe the susceptible-infected-susceptible (SIS) model with higher-order interactions, where, being the same with the traditional SIS model, a node can take either susceptible (S) state or infected (I) state at a time, whereas the infection mechanism is extended with higher-order interactions. As illustrated in Fig.~\ref{fig1}(b), a susceptible node in a simplex can be infected by its infected nodes through edges (traditional pairwise interaction) and can be also infected by higher-order ways (group interactions) through simplices, which is different from the traditional case that a susceptible node can be only infected by one of its infected neighbors through edges. Note a susceptible node can get simplicial infection only when all the rest neighbors in the same simplex are infected. Besides, the order compatibility of simplicial complexes makes it possible for a node to be infected through lower-order interaction, yet in a higher-order simplex. Finally, due to the additional interaction forms, we also need more parameters to describe the spreading dynamics. We assign each interaction form a unique parameter as the infectious rate because they are actually possible to have different values. Specifically, we use $\beta_1, \beta_2, ..., \beta_p$ to describe the infectious rate of 1-simplices (edges), 2-simplices (colored triangles), ..., $p$-simplices, respectively. As for the recovery process, which is a spontaneous process with no interactions, we use $\mu$ to describe the recovery rate of all the nodes for simplicity.

In order to see clearly the evolutionary process of the simplicial SIS model, we first visualize the result on a small network (including higher-order links). In Fig.~\ref{fig2}, we compare the result of simplicial spreading to the traditional SIS process. For simplicity, we only consider interactions up to second order (i.e., interaction through 2-simplices), and we use $\lambda_1$ and $\lambda_2$ (normalized infectious rate) to describe pairwise and higher-order interactions, respectively.

We know that in the traditional SIS process, if the infectious rate $\lambda_1$ becomes larger than a certain threshold, the infection density (portion of infected individuals in the network) will gradually increase and finally become stable at a certain value between 0 and 1. As shown in Fig.~\ref{fig2}(a), after the initial infection at time 0, the number of infected individuals gradually increases until time 200 and then reaches a stable state. As for Fig.~\ref{fig2}(b) when $\lambda_2>0$, there is no significant difference between the two processes at early time. However, one can see that after time 100, the infection density starts increasing much faster than the traditional process, and the stable infection density is also significantly larger than that of the traditional SIS model that only allows pairwise interactions. Obviously, at initial times, due to the presence of only a small number of infected individuals, the probability that two neighbors of a given node are simultaneously infected is extremely low, and therefore only a negligible number of 2-simplices are getting activated. As the number of infected individuals increases, more and more 2-simplices become activated, which makes higher-order interactions start playing an important role. It is worth noticing that for a SIS process with limited initial infected individuals, it is difficult or almost impossible for the infection to gain prevalence if only higher-order interactions are allowed. In other words, pairwise interactions are needed to activate  higher-order structures and after that higher-order interactions takes the lead on the overall process, as they will remarkably accelerate the infection process and make more individuals get infected at the stable state.

\begin{figure}[tb]
\centering
\includegraphics[width=\linewidth, scale=1.00]{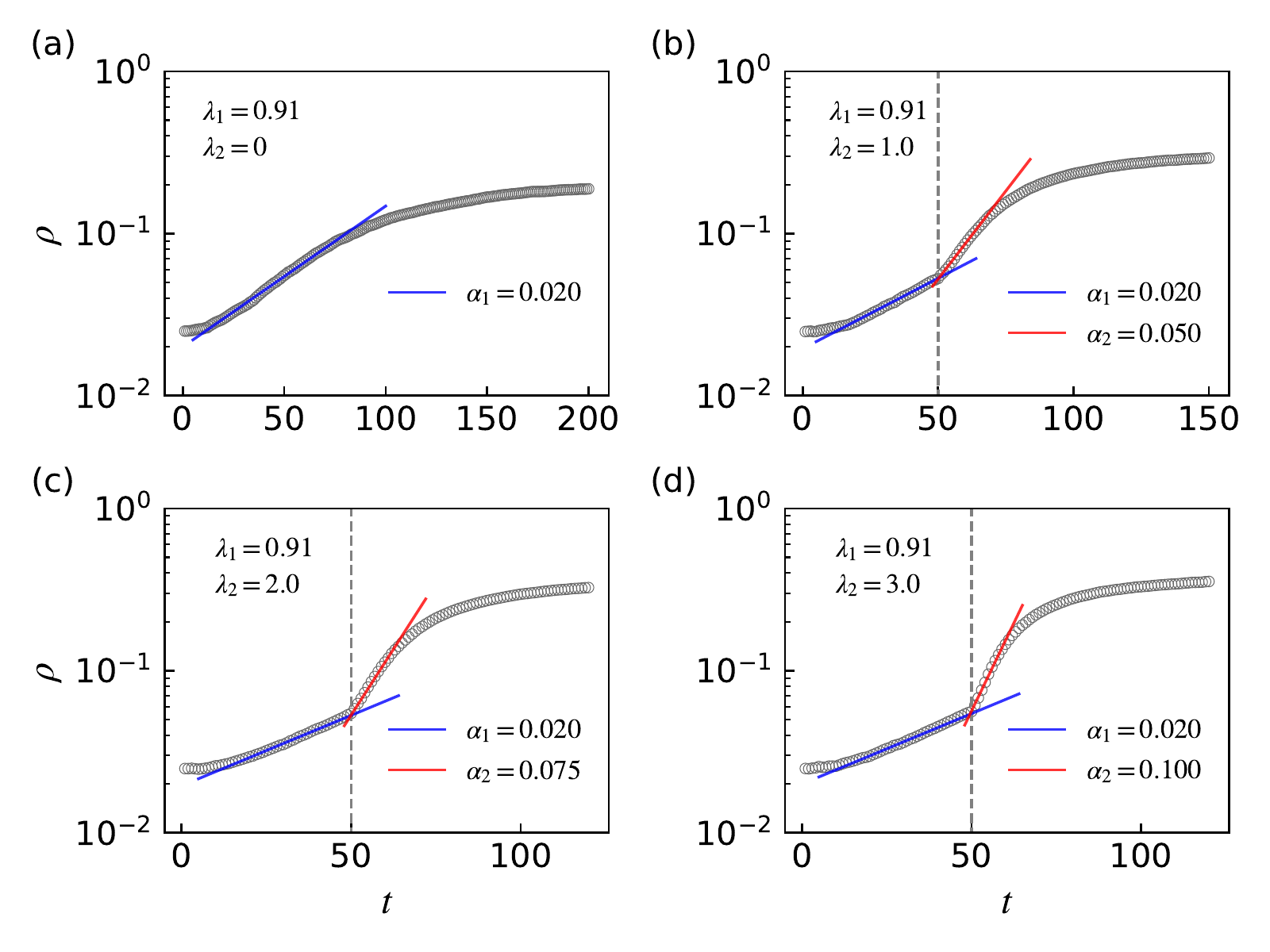}
\caption{\label{fig3}The infection density $\rho$ of the SIS process as a function of time $t$ on a synthetic network. The network has 1,000 nodes, 4,140 edges, and 1,401 triangles. Grey dashed line marks the time at which second-order interaction kicks in. Blue and red lines are some asymptotic trend curves, which subject to exponential functions with the exponents being $\alpha_1$ and $\alpha_2$, respectively. In the simulations, we take 85\% randomly chosen triangles as 2-simplices. All the results are obtained by taking average over 1,000 runs.}
\end{figure}

To quantitatively demonstrate the acceleration and promotion of the infection, we investigate the infection density $\rho$ of the simplicial SIS model on a large synthetic network, made of $N=1,000$ nodes, $4,140$ 1-simplices (edges) and $1,401$ 2-simplices, generated by the extended Barabási Albert model introduced in Ref. \cite{growing}. In our simulations, we allow higher-order interactions only after a certain number of 2-simplices being activated. For simplicity, in our case, we take $t=50$ as the initial time for switching on triadwise interactions (at that time, indeed, there are on average 5\% activated 2-simplices). Since the evolution process of the SIS model give rise to a logistic-like curve and the increasing part of it can be approximately considered as an  exponential, in Fig.~\ref{fig3} we also show some asymptotic trend curves that obey exponential functions with certain exponents. As compared to Fig.~\ref{fig3}(a), in Fig.~\ref{fig3}(b)$\sim$(d) one clearly see that after $t=50$ (i.e. after having activated the higher-order interactions), both the exponents of the curves and the final infection density become larger, which implies that higher-order interactions lead to a faster increase of the infection density $\rho$ and to a larger prevalence of infected individuals in the stable state. In addition, one can see that the larger the second-order infectious rate $\lambda_2$ is, the faster  $\rho$ increases and the more individuals get infected in the stable state.

\begin{figure}[htb]
\centering
\includegraphics[width=\linewidth, scale=1.00]{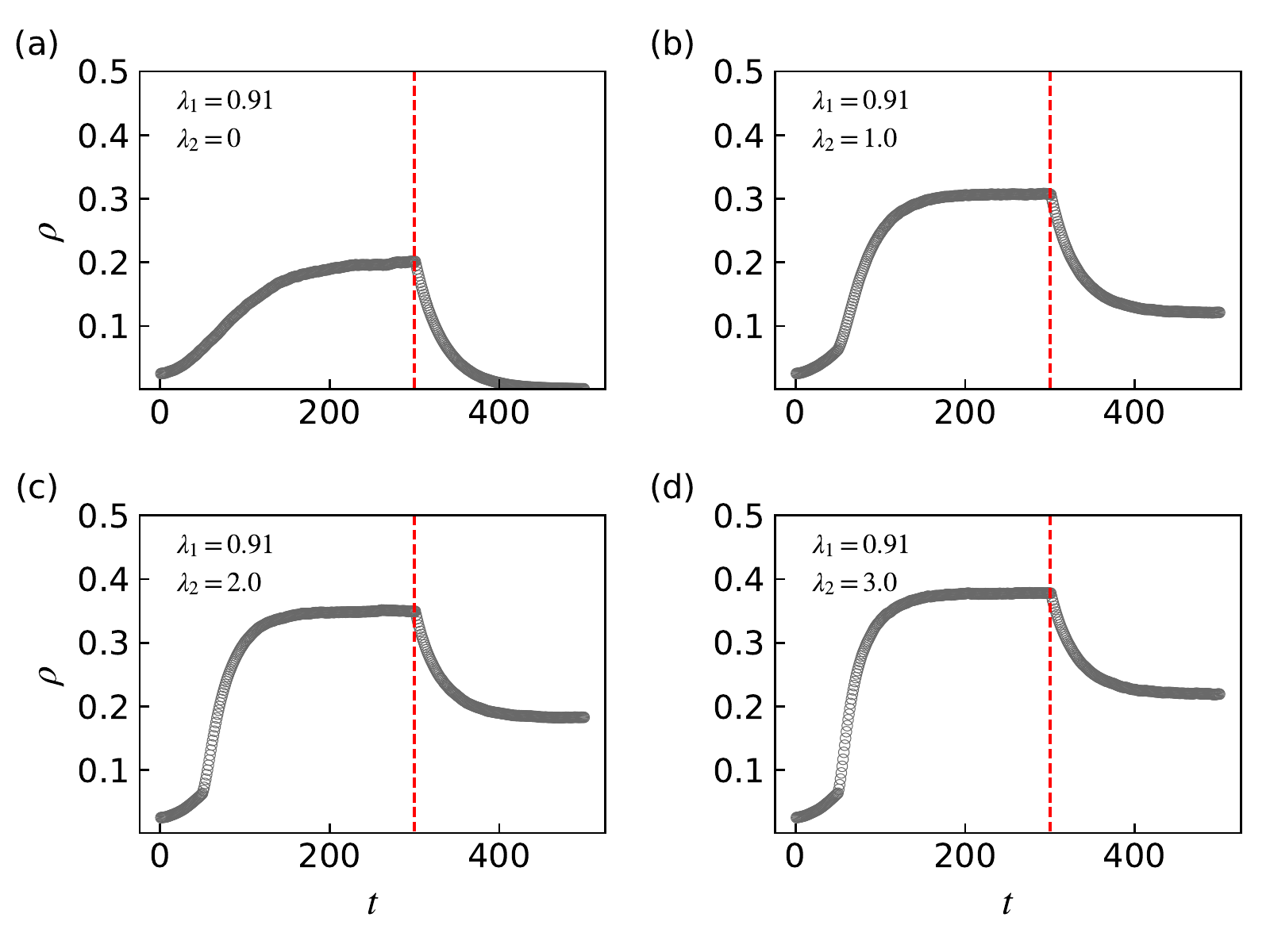}
\caption{\label{fig4}The infection density $\rho$ of the SIS process as a function of time $t$ on a synthetic network (see main text for the specification of the network). The experiment configuration is the same as that in Fig. 3. Red dashed lines denote where we shut down pairwise interactions.}
\end{figure}

As mentioned above, higher-order interactions are not enough to ignite a spreading process from an initial state that is disease-free or have just a few individuals infected. Now we move to show that mere higher-order interactions are instead able to sustain the contagion in a population if there is already a certain amount of infected individuals. Then, we make more simulations with the same configurations considered in Fig.~\ref{fig3}. This time, we shut down the pairwise (first-order) interactions and only allow second-order ones after the infection density becomes stable. Remarkably, as shown in Fig.~\ref{fig4}, after shutting down the pairwise interaction, the infection density $\rho$ will immediately decrease, but if $\lambda_2$ is sufficiently large, it will gradually become stable at another smaller point instead of dying out. This means that a mere higher-order interaction is able to maintain a spreading process if there is sufficient infected individuals and the higher-order infectious rate is large enough.

\begin{figure*}
\centering
\includegraphics[width=\linewidth, scale=1.00]{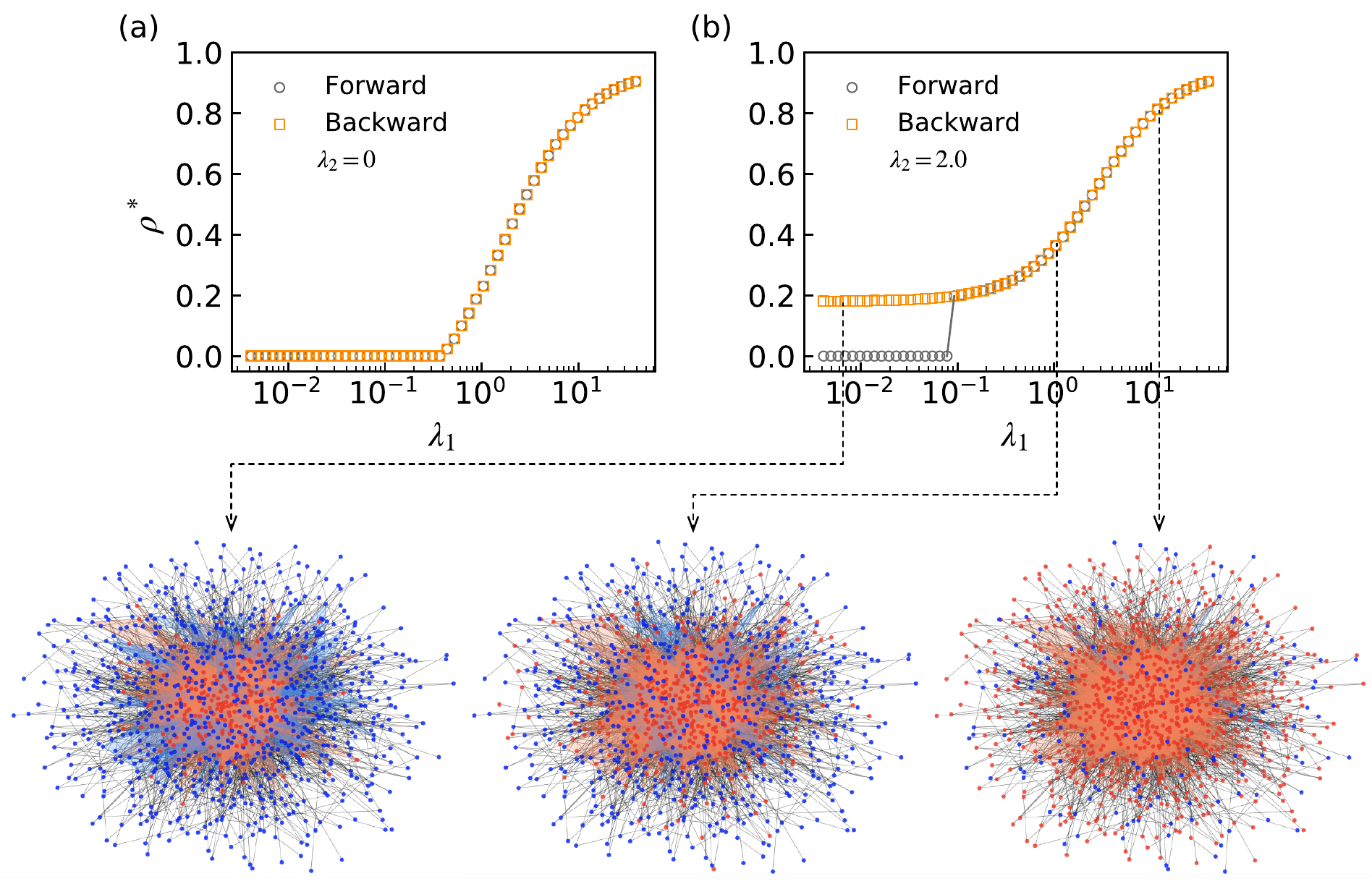}
\caption{\label{fig5} The final infection density $\rho^*$ as a function of infectious rate $\lambda_1$ and $\lambda_2$. (a) and (b) are the results of the traditional and simplicial SIS model, respectively. Forward processes denote there are few (less than 3\%) infected individuals at the initial time, and backward processes mean a large amount (more than 80\%). The three network snapshots demonstrate the final conditions corresponding to the three selected point. The network configuration is the same as Fig. 4. }
\end{figure*}


In addition to the dynamical process, we also study the evolution of the final infection density. For the traditional SIS model, the final infection density $\rho^*$ (the value of $\rho$ at the stable state) is mainly determined by the (normalized) infectious rate $\lambda_1$. As shown in Fig.~\ref{fig5}(a), $\rho^*$ changes with $\lambda_1$ through a continuous transition, and there is no difference between forward and backward processes, which means that the final infection density does not depend upon the specific initial condition. 

In Fig. 5(b), we report instead the phase transition of $\rho^*$ under the simplicial SIS model, and some new phenomena appear. First, the outbreak threshold for the forward processes becomes much smaller under the influence of second-order interactions. Second, the final infection density $\rho^*$ experiences an abrupt, discontinuous, phase transition. This means that spreadings are easier to break out with higher-order interactions,  and become stable at a relatively larger point. More importantly one can see that, for backward processes, even if $\lambda_1$ is almost negligible, the final infection density $\rho^*$ is still non-zero.  This is consistent with what already discussed concerning the fact that, when there exist sufficient higher-order interactions, once a spreading process is activated, the infection will persist even if pairwise interactions are disabled. We also draw three snapshots of the network in Fig.~\ref{fig5},  to illustrate the final condition more clearly. One indeed clearly sees in the first snapshot that infected individuals only exist in 2-simplices (in the final condition and for a nearly vanishing $\lambda_1$), which means that second-order interactions are the ones that effectively sustain the infection. On the opposite, the other two snapshots refer to cases where there are both pairwise and second-order interactions, and one can clearly see that  infected individuals exist in all kinds of sub-structures. Once again, more activated 2-simplices occur when $\lambda_1$ becomes larger, which indicates that a larger $\lambda_1$ is able to activate more higher-order interactions, and those latter ones facilitate in return the infection process.

The new phenomena described so far are not peculiar of the specific network configurations that we have used in our simulations. One, indeed, can easily find a generalized law of the second-order SIS model by mean-field methods (which have already been extended to higher-order cases in previous studies \cite{iacopini2019simplicial}). According to the homogeneous mixing hypothesis, one can describe the dynamical evolution of the infection density $\rho$ as a differential equation:

\begin{equation}
\label{eq1}
	\dot{\rho}=-\rho+\lambda_1\rho(1-\rho)+\lambda_2\rho^2(1-\rho)
\end{equation}

Letting $\dot{\rho}=0$, one can solve Eq. (\ref{eq1}), and obtain the final infection density $\rho^*$. The equation can be rewritten as:

\begin{equation}
\label{eq2}
\begin{split}
	& \rho-\lambda_1\rho(1-\rho)-\lambda_2\rho^2(1-\rho)=0\\
	\Rightarrow & \lambda_2\rho^2+(\lambda_1-\lambda_2)\rho +1-\lambda_1=0, 
\end{split}
\end{equation}

and one can obtain $\rho^*$ as:

\begin{equation}
	\label{eq3}
	\rho_{\pm}^*=\frac{\lambda_2-\lambda_1\pm\sqrt{(\lambda_1-\lambda_2)^2-4\lambda_2(1-\lambda_1)}}{2\lambda_2}.
\end{equation}

\begin{figure}[tb]
	\centering
	\includegraphics[width=\linewidth, scale=1.00]{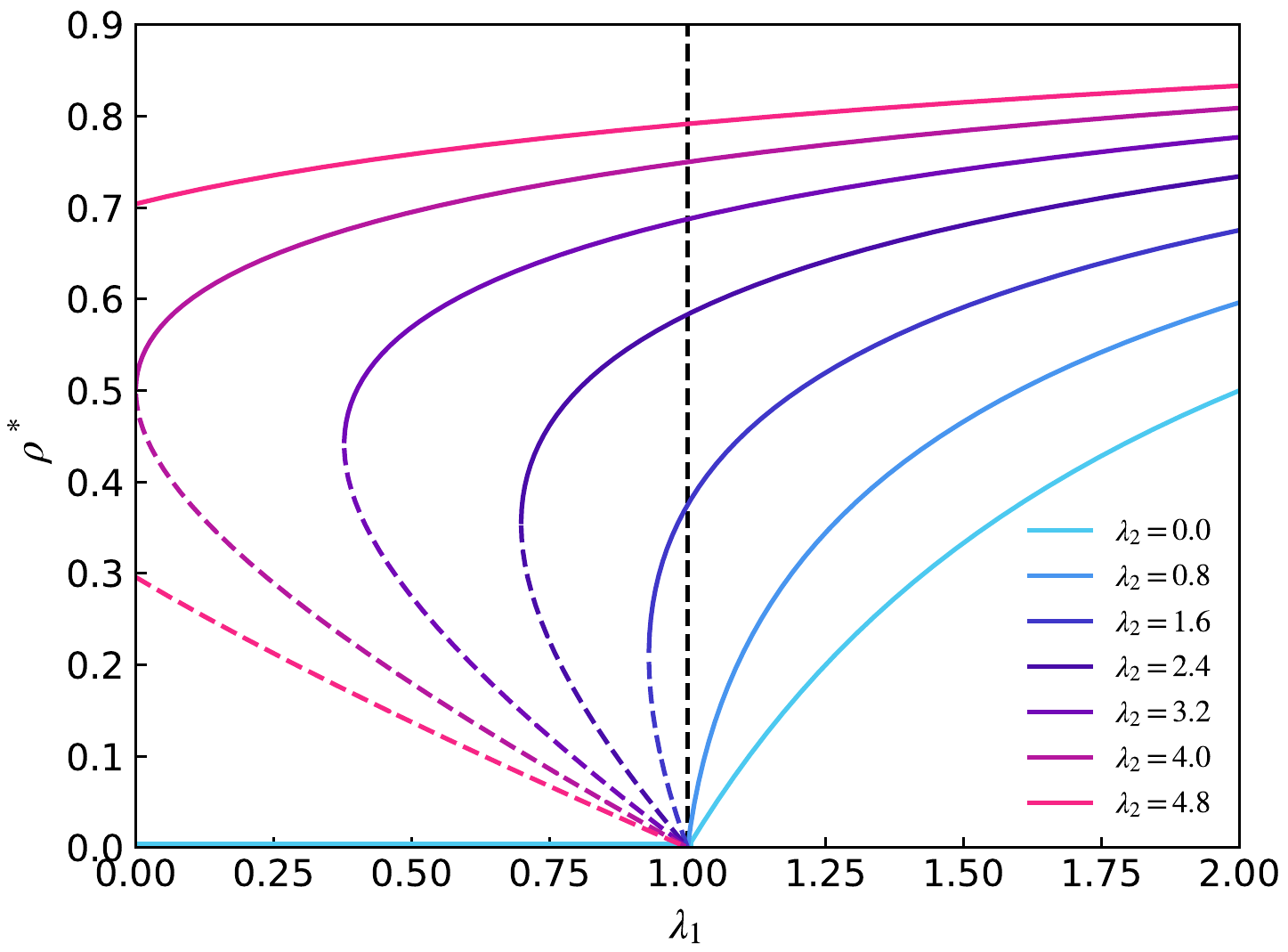}
	\caption{\label{fig6} The mean-field solutions of the final infection density $\rho^*$. Solid and dashed curves denote $\rho^*_{+}$ and $\rho^*_{-}$, respectively.}
\end{figure}

In Fig.~\ref{fig6}, we report the solutions of $\rho^*$ for a few parameter values of $\lambda_1$ and $\lambda_2$. It is obvious that under certain conditions, both $\rho^*_{+}$ and $\rho^*_{-}$ are positive. Now, if one rewrites Eq.~(\ref{eq1}) as:
\begin{equation}
\label{eq4}
	\dot{\rho}=-\rho(\rho-\rho^*_{+})(\rho-\rho^*_{-}),
\end{equation}
one can find that $\rho^*_{-}$ is an unstable solution, which determines a bi-stable regime, because any perturbation would make $\rho$ converge to either $\rho^*_{+}$ or zero, but it will never relax back to $\rho^*_{-}$. Thus, for second-order SIS model with a relatively large $\lambda_2$, the outbreak is activated once there are more than $\rho^*_{-}$ infected individuals, which is different from the traditional case where outbreaks happen only when $\lambda_1>1$. This also explains why  the outbreak threshold becomes smaller with second-order interactions, and $\rho^*$ changes through a discontinuous way. It is worth mentioning that as $\lambda_2$ increases, the bi-stable region also expands. More importantly, when $\lambda_2>4$, the bi-stable region is able to cover the entire interval starting from $\lambda_1=0$, which theoretically demonstrates the possibility that higher-order interactions can maintain an activated spreading process on their own.

In summary, we investigated the effects of higher-order interactions on the social contagion and its dynamical process. We find that at the initial phase of a spreading process, higher-order interaction makes no obvious influence. However, as there are sufficient higher-order structures being activated, higher-order interaction will then significantly accelerate the spreading process. Also, the final infection density at the stable state will be much higher than that of the traditional model. This may indicate that in reality, the spreading of some new ideas or forms in the population might turn out to be faster than we estimate with the traditional models that only consider pairwise interactions.

In some ways,  higher-order interactions act very similarly to peer pressure mechanisms, consisting in the fact that people tend to make the same actions as their friends or accept ideas that their friends adopt. Apparently, we cannot omit the influence of this kind of peer pressure especially when analyzing social contagion problems. As we demonstrate both in our simulation and theoretical results, under certain conditions, the second-order interaction could maintain a spreading process without pairwise interactions. This might be somewhat inspiring for products marketing or new concepts spreading. For example, companies may be able to cut down the cost for marketing their products once the users exceed a certain amount because the peer pressure (if it is high enough) will then maintain at least part of the users.

\section{Acknowledgements}
This research was supported by the Israel Science Foundation (grant No. 499/19) and by the US National Science Foundation-CRISP award no. 1735505.


\begin{thebibliography}{33}%
	\makeatletter
	\providecommand \@ifxundefined [1]{%
		\@ifx{#1\undefined}
	}%
	\providecommand \@ifnum [1]{%
		\ifnum #1\expandafter \@firstoftwo
		\else \expandafter \@secondoftwo
		\fi
	}%
	\providecommand \@ifx [1]{%
		\ifx #1\expandafter \@firstoftwo
		\else \expandafter \@secondoftwo
		\fi
	}%
	\providecommand \natexlab [1]{#1}%
	\providecommand \enquote  [1]{``#1''}%
	\providecommand \bibnamefont  [1]{#1}%
	\providecommand \bibfnamefont [1]{#1}%
	\providecommand \citenamefont [1]{#1}%
	\providecommand \href@noop [0]{\@secondoftwo}%
	\providecommand \href [0]{\begingroup \@sanitize@url \@href}%
	\providecommand \@href[1]{\@@startlink{#1}\@@href}%
	\providecommand \@@href[1]{\endgroup#1\@@endlink}%
	\providecommand \@sanitize@url [0]{\catcode `\\12\catcode `\$12\catcode
		`\&12\catcode `\#12\catcode `\^12\catcode `\_12\catcode `\%12\relax}%
	\providecommand \@@startlink[1]{}%
	\providecommand \@@endlink[0]{}%
	\providecommand \url  [0]{\begingroup\@sanitize@url \@url }%
	\providecommand \@url [1]{\endgroup\@href {#1}{\urlprefix }}%
	\providecommand \urlprefix  [0]{URL }%
	\providecommand \Eprint [0]{\href }%
	\providecommand \doibase [0]{https://doi.org/}%
	\providecommand \selectlanguage [0]{\@gobble}%
	\providecommand \bibinfo  [0]{\@secondoftwo}%
	\providecommand \bibfield  [0]{\@secondoftwo}%
	\providecommand \translation [1]{[#1]}%
	\providecommand \BibitemOpen [0]{}%
	\providecommand \bibitemStop [0]{}%
	\providecommand \bibitemNoStop [0]{.\EOS\space}%
	\providecommand \EOS [0]{\spacefactor3000\relax}%
	\providecommand \BibitemShut  [1]{\csname bibitem#1\endcsname}%
	\let\auto@bib@innerbib\@empty
	\bibitem [{\citenamefont {Barr\'{a}t}\ \emph {et~al.}(2008)\citenamefont
		{Barr\'{a}t}, \citenamefont {Barth\'{e}lemy},\ and\ \citenamefont
		{Vespignani}}]{barrat2008dynamical}%
	\BibitemOpen
	\bibfield  {author} {\bibinfo {author} {\bibfnamefont {A.}~\bibnamefont
			{Barr\'{a}t}}, \bibinfo {author} {\bibfnamefont {M.}~\bibnamefont
			{Barth\'{e}lemy}},\ and\ \bibinfo {author} {\bibfnamefont {A.}~\bibnamefont
			{Vespignani}},\ }\href@noop {} {\emph {\bibinfo {title} {Dynamical processes
				on complex networks}}}\ (\bibinfo  {publisher} {Cambridge University Press},\
	\bibinfo {year} {2008})\BibitemShut {NoStop}%
	\bibitem [{\citenamefont {Pastor-Satorras}\ \emph {et~al.}(2015)\citenamefont
		{Pastor-Satorras}, \citenamefont {Castellano}, \citenamefont {Van~Mieghem},\
		and\ \citenamefont {Vespignani}}]{pastor2015epidemic}%
	\BibitemOpen
	\bibfield  {author} {\bibinfo {author} {\bibfnamefont {R.}~\bibnamefont
			{Pastor-Satorras}}, \bibinfo {author} {\bibfnamefont {C.}~\bibnamefont
			{Castellano}}, \bibinfo {author} {\bibfnamefont {P.}~\bibnamefont
			{Van~Mieghem}},\ and\ \bibinfo {author} {\bibfnamefont {A.}~\bibnamefont
			{Vespignani}},\ }\bibfield  {title} {\bibinfo {title} {Epidemic processes in
			complex networks},\ }\href@noop {} {\bibfield  {journal} {\bibinfo  {journal}
			{Reviews of Modern Physics}\ }\textbf {\bibinfo {volume} {87}},\ \bibinfo
		{pages} {925} (\bibinfo {year} {2015})}\BibitemShut {NoStop}%
	\bibitem [{\citenamefont {Pastor-Satorras}\ and\ \citenamefont
		{Vespignani}(2001)}]{pastor2001epidemic}%
	\BibitemOpen
	\bibfield  {author} {\bibinfo {author} {\bibfnamefont {R.}~\bibnamefont
			{Pastor-Satorras}}\ and\ \bibinfo {author} {\bibfnamefont {A.}~\bibnamefont
			{Vespignani}},\ }\bibfield  {title} {\bibinfo {title} {Epidemic spreading in
			scale-free networks},\ }\href@noop {} {\bibfield  {journal} {\bibinfo
			{journal} {Physical Review Letters}\ }\textbf {\bibinfo {volume} {86}},\
		\bibinfo {pages} {3200} (\bibinfo {year} {2001})}\BibitemShut {NoStop}%
	\bibitem [{\citenamefont {Bogu\~{n}{\'a}}\ and\ \citenamefont
		{Pastor-Satorras}(2002)}]{boguna2002epidemic}%
	\BibitemOpen
	\bibfield  {author} {\bibinfo {author} {\bibfnamefont {M.}~\bibnamefont
			{Bogu\~{n}{\'a}}}\ and\ \bibinfo {author} {\bibfnamefont {R.}~\bibnamefont
			{Pastor-Satorras}},\ }\bibfield  {title} {\bibinfo {title} {Epidemic
			spreading in correlated complex networks},\ }\href@noop {} {\bibfield
		{journal} {\bibinfo  {journal} {Physical Review E}\ }\textbf {\bibinfo
			{volume} {66}},\ \bibinfo {pages} {047104} (\bibinfo {year}
		{2002})}\BibitemShut {NoStop}%
	\bibitem [{\citenamefont {Boccaletti}\ \emph {et~al.}(2006)\citenamefont
		{Boccaletti}, \citenamefont {Latora}, \citenamefont {Moreno}, \citenamefont
		{Chavez},\ and\ \citenamefont {Hwang}}]{boccaletti2006complex}%
	\BibitemOpen
	\bibfield  {author} {\bibinfo {author} {\bibfnamefont {S.}~\bibnamefont
			{Boccaletti}}, \bibinfo {author} {\bibfnamefont {V.}~\bibnamefont {Latora}},
		\bibinfo {author} {\bibfnamefont {Y.}~\bibnamefont {Moreno}}, \bibinfo
		{author} {\bibfnamefont {M.}~\bibnamefont {Chavez}},\ and\ \bibinfo {author}
		{\bibfnamefont {D.-U.}\ \bibnamefont {Hwang}},\ }\bibfield  {title} {\bibinfo
		{title} {Complex networks: Structure and dynamics},\ }\href@noop {}
	{\bibfield  {journal} {\bibinfo  {journal} {Physics Reports}\ }\textbf
		{\bibinfo {volume} {424}},\ \bibinfo {pages} {175} (\bibinfo {year}
		{2006})}\BibitemShut {NoStop}%
	\bibitem [{\citenamefont {Watts}(2011)}]{watts2011simple}%
	\BibitemOpen
	\bibfield  {author} {\bibinfo {author} {\bibfnamefont {D.~J.}\ \bibnamefont
			{Watts}},\ }\bibfield  {title} {\bibinfo {title} {A simple model of global
			cascades on random networks},\ }in\ \href@noop {} {\emph {\bibinfo
			{booktitle} {The Structure and Dynamics of Networks}}}\ (\bibinfo
	{publisher} {Princeton University Press},\ \bibinfo {year} {2011})\ pp.\
	\bibinfo {pages} {497--502}\BibitemShut {NoStop}%
	\bibitem [{\citenamefont {Simonsen}\ \emph {et~al.}(2008)\citenamefont
		{Simonsen}, \citenamefont {Buzna}, \citenamefont {Peters}, \citenamefont
		{Bornholdt},\ and\ \citenamefont {Helbing}}]{simonsen2008transient}%
	\BibitemOpen
	\bibfield  {author} {\bibinfo {author} {\bibfnamefont {I.}~\bibnamefont
			{Simonsen}}, \bibinfo {author} {\bibfnamefont {L.}~\bibnamefont {Buzna}},
		\bibinfo {author} {\bibfnamefont {K.}~\bibnamefont {Peters}}, \bibinfo
		{author} {\bibfnamefont {S.}~\bibnamefont {Bornholdt}},\ and\ \bibinfo
		{author} {\bibfnamefont {D.}~\bibnamefont {Helbing}},\ }\bibfield  {title}
	{\bibinfo {title} {Transient dynamics increasing network vulnerability to
			cascading failures},\ }\href@noop {} {\bibfield  {journal} {\bibinfo
			{journal} {Physical Review Letters}\ }\textbf {\bibinfo {volume} {100}},\
		\bibinfo {pages} {218701} (\bibinfo {year} {2008})}\BibitemShut {NoStop}%
	\bibitem [{\citenamefont {Newman}(2018)}]{newman2018networks}%
	\BibitemOpen
	\bibfield  {author} {\bibinfo {author} {\bibfnamefont {M.~E.}\ \bibnamefont
			{Newman}},\ }\href@noop {} {\emph {\bibinfo {title} {Networks: An
				introduction}}}\ (\bibinfo  {publisher} {Oxford University Press},\ \bibinfo
	{year} {2018})\BibitemShut {NoStop}%
	\bibitem [{\citenamefont {Newman}(2003)}]{newman2003structure}%
	\BibitemOpen
	\bibfield  {author} {\bibinfo {author} {\bibfnamefont {M.~E.}\ \bibnamefont
			{Newman}},\ }\bibfield  {title} {\bibinfo {title} {The structure and function
			of complex networks},\ }\href@noop {} {\bibfield  {journal} {\bibinfo
			{journal} {SIAM Review}\ }\textbf {\bibinfo {volume} {45}},\ \bibinfo {pages}
		{167} (\bibinfo {year} {2003})}\BibitemShut {NoStop}%
	\bibitem [{\citenamefont {Wu}\ \emph {et~al.}(2015)\citenamefont {Wu},
		\citenamefont {Menichetti}, \citenamefont {Rahmede},\ and\ \citenamefont
		{Bianconi}}]{wu2015emergent}%
	\BibitemOpen
	\bibfield  {author} {\bibinfo {author} {\bibfnamefont {Z.}~\bibnamefont
			{Wu}}, \bibinfo {author} {\bibfnamefont {G.}~\bibnamefont {Menichetti}},
		\bibinfo {author} {\bibfnamefont {C.}~\bibnamefont {Rahmede}},\ and\ \bibinfo
		{author} {\bibfnamefont {G.}~\bibnamefont {Bianconi}},\ }\bibfield  {title}
	{\bibinfo {title} {Emergent complex network geometry},\ }\href@noop {}
	{\bibfield  {journal} {\bibinfo  {journal} {Scientific Reports}\ }\textbf
		{\bibinfo {volume} {5}},\ \bibinfo {pages} {1} (\bibinfo {year}
		{2015})}\BibitemShut {NoStop}%
	\bibitem [{\citenamefont {Cowan}\ and\ \citenamefont
		{Jonard}(2004)}]{cowan2004network}%
	\BibitemOpen
	\bibfield  {author} {\bibinfo {author} {\bibfnamefont {R.}~\bibnamefont
			{Cowan}}\ and\ \bibinfo {author} {\bibfnamefont {N.}~\bibnamefont {Jonard}},\
	}\bibfield  {title} {\bibinfo {title} {Network structure and the diffusion of
			knowledge},\ }\href@noop {} {\bibfield  {journal} {\bibinfo  {journal}
			{Journal of Economic Dynamics and Control}\ }\textbf {\bibinfo {volume}
			{28}},\ \bibinfo {pages} {1557} (\bibinfo {year} {2004})}\BibitemShut
	{NoStop}%
	\bibitem [{\citenamefont {Hens}\ \emph {et~al.}(2019)\citenamefont {Hens},
		\citenamefont {Harush}, \citenamefont {Haber}, \citenamefont {Cohen},\ and\
		\citenamefont {Barzel}}]{Hens2019network}%
	\BibitemOpen
	\bibfield  {author} {\bibinfo {author} {\bibfnamefont {C.}~\bibnamefont
			{Hens}}, \bibinfo {author} {\bibfnamefont {U.}~\bibnamefont {Harush}},
		\bibinfo {author} {\bibfnamefont {S.}~\bibnamefont {Haber}}, \bibinfo
		{author} {\bibfnamefont {R.}~\bibnamefont {Cohen}},\ and\ \bibinfo {author}
		{\bibfnamefont {B.}~\bibnamefont {Barzel}},\ }\bibfield  {title} {\bibinfo
		{title} {Spatiotemporal signal propagation in complex networks},\ }\href@noop
	{} {\bibfield  {journal} {\bibinfo  {journal} {Nature Physics}\ }\textbf
		{\bibinfo {volume} {15}},\ \bibinfo {pages} {403} (\bibinfo {year}
		{2019})}\BibitemShut {NoStop}%
	\bibitem [{\citenamefont {Harush}\ and\ \citenamefont
		{Barzel}(2017)}]{Harush2017Patterns}%
	\BibitemOpen
	\bibfield  {author} {\bibinfo {author} {\bibfnamefont {U.}~\bibnamefont
			{Harush}}\ and\ \bibinfo {author} {\bibfnamefont {B.}~\bibnamefont
			{Barzel}},\ }\bibfield  {title} {\bibinfo {title} {Dynamic patterns of
			information flow in complex networks},\ }\href@noop {} {\bibfield  {journal}
		{\bibinfo  {journal} {Nature Communications}\ }\textbf {\bibinfo {volume}
			{8}},\ \bibinfo {pages} {2181} (\bibinfo {year} {2017})}\BibitemShut
	{NoStop}%
	\bibitem [{\citenamefont {Centola}\ and\ \citenamefont
		{Macy}(2007)}]{centola2007complex}%
	\BibitemOpen
	\bibfield  {author} {\bibinfo {author} {\bibfnamefont {D.}~\bibnamefont
			{Centola}}\ and\ \bibinfo {author} {\bibfnamefont {M.}~\bibnamefont {Macy}},\
	}\bibfield  {title} {\bibinfo {title} {Complex contagions and the weakness of
			long ties},\ }\href@noop {} {\bibfield  {journal} {\bibinfo  {journal}
			{American Journal of Sociology}\ }\textbf {\bibinfo {volume} {113}},\
		\bibinfo {pages} {702} (\bibinfo {year} {2007})}\BibitemShut {NoStop}%
	\bibitem [{\citenamefont {Benson}\ \emph {et~al.}(2016)\citenamefont {Benson},
		\citenamefont {Gleich},\ and\ \citenamefont {Leskovec}}]{benson2016higher}%
	\BibitemOpen
	\bibfield  {author} {\bibinfo {author} {\bibfnamefont {A.~R.}\ \bibnamefont
			{Benson}}, \bibinfo {author} {\bibfnamefont {D.~F.}\ \bibnamefont {Gleich}},\
		and\ \bibinfo {author} {\bibfnamefont {J.}~\bibnamefont {Leskovec}},\
	}\bibfield  {title} {\bibinfo {title} {Higher-order organization of complex
			networks},\ }\href@noop {} {\bibfield  {journal} {\bibinfo  {journal}
			{Science}\ }\textbf {\bibinfo {volume} {353}},\ \bibinfo {pages} {163}
		(\bibinfo {year} {2016})}\BibitemShut {NoStop}%
	\bibitem [{\citenamefont {Grilli}\ \emph {et~al.}(2017)\citenamefont {Grilli},
		\citenamefont {Barab{\'a}s}, \citenamefont {Michalska-Smith},\ and\
		\citenamefont {Allesina}}]{grilli2017higher}%
	\BibitemOpen
	\bibfield  {author} {\bibinfo {author} {\bibfnamefont {J.}~\bibnamefont
			{Grilli}}, \bibinfo {author} {\bibfnamefont {G.}~\bibnamefont {Barab{\'a}s}},
		\bibinfo {author} {\bibfnamefont {M.~J.}\ \bibnamefont {Michalska-Smith}},\
		and\ \bibinfo {author} {\bibfnamefont {S.}~\bibnamefont {Allesina}},\
	}\bibfield  {title} {\bibinfo {title} {Higher-order interactions stabilize
			dynamics in competitive network models},\ }\href@noop {} {\bibfield
		{journal} {\bibinfo  {journal} {Nature}\ }\textbf {\bibinfo {volume} {548}},\
		\bibinfo {pages} {210} (\bibinfo {year} {2017})}\BibitemShut {NoStop}%
	\bibitem [{\citenamefont {Guilbeault}\ \emph {et~al.}(2018)\citenamefont
		{Guilbeault}, \citenamefont {Becker},\ and\ \citenamefont
		{Centola}}]{guilbeault2018complex}%
	\BibitemOpen
	\bibfield  {author} {\bibinfo {author} {\bibfnamefont {D.}~\bibnamefont
			{Guilbeault}}, \bibinfo {author} {\bibfnamefont {J.}~\bibnamefont {Becker}},\
		and\ \bibinfo {author} {\bibfnamefont {D.}~\bibnamefont {Centola}},\
	}\bibfield  {title} {\bibinfo {title} {Complex contagions: A decade in
			review},\ }in\ \href@noop {} {\emph {\bibinfo {booktitle} {Complex Spreading
				Phenomena in Social Systems}}}\ (\bibinfo  {publisher} {Springer},\ \bibinfo
	{year} {2018})\ pp.\ \bibinfo {pages} {3--25}\BibitemShut {NoStop}%
	\bibitem [{\citenamefont {Centola}(2010)}]{centola2010spread}%
	\BibitemOpen
	\bibfield  {author} {\bibinfo {author} {\bibfnamefont {D.}~\bibnamefont
			{Centola}},\ }\bibfield  {title} {\bibinfo {title} {The spread of behavior in
			an online social network experiment},\ }\href@noop {} {\bibfield  {journal}
		{\bibinfo  {journal} {Science}\ }\textbf {\bibinfo {volume} {329}},\ \bibinfo
		{pages} {1194} (\bibinfo {year} {2010})}\BibitemShut {NoStop}%
	\bibitem [{\citenamefont {Ugander}\ \emph {et~al.}(2012)\citenamefont
		{Ugander}, \citenamefont {Backstrom}, \citenamefont {Marlow},\ and\
		\citenamefont {Kleinberg}}]{ugander2012structural}%
	\BibitemOpen
	\bibfield  {author} {\bibinfo {author} {\bibfnamefont {J.}~\bibnamefont
			{Ugander}}, \bibinfo {author} {\bibfnamefont {L.}~\bibnamefont {Backstrom}},
		\bibinfo {author} {\bibfnamefont {C.}~\bibnamefont {Marlow}},\ and\ \bibinfo
		{author} {\bibfnamefont {J.}~\bibnamefont {Kleinberg}},\ }\bibfield  {title}
	{\bibinfo {title} {Structural diversity in social contagion},\ }\href@noop {}
	{\bibfield  {journal} {\bibinfo  {journal} {Proceedings of the National
				Academy of Sciences}\ }\textbf {\bibinfo {volume} {109}},\ \bibinfo {pages}
		{5962} (\bibinfo {year} {2012})}\BibitemShut {NoStop}%
	\bibitem [{\citenamefont {Chan}\ \emph {et~al.}(2013)\citenamefont {Chan},
		\citenamefont {Carlsson},\ and\ \citenamefont {Rabadan}}]{chan2013topology}%
	\BibitemOpen
	\bibfield  {author} {\bibinfo {author} {\bibfnamefont {J.~M.}\ \bibnamefont
			{Chan}}, \bibinfo {author} {\bibfnamefont {G.}~\bibnamefont {Carlsson}},\
		and\ \bibinfo {author} {\bibfnamefont {R.}~\bibnamefont {Rabadan}},\
	}\bibfield  {title} {\bibinfo {title} {Topology of viral evolution},\
	}\href@noop {} {\bibfield  {journal} {\bibinfo  {journal} {Proceedings of the
				National Academy of Sciences}\ }\textbf {\bibinfo {volume} {110}},\ \bibinfo
		{pages} {18566} (\bibinfo {year} {2013})}\BibitemShut {NoStop}%
	\bibitem [{\citenamefont {Nanda}\ and\ \citenamefont
		{Sazdanovi{\'c}}(2014)}]{nanda2014simplicial}%
	\BibitemOpen
	\bibfield  {author} {\bibinfo {author} {\bibfnamefont {V.}~\bibnamefont
			{Nanda}}\ and\ \bibinfo {author} {\bibfnamefont {R.}~\bibnamefont
			{Sazdanovi{\'c}}},\ }\bibfield  {title} {\bibinfo {title} {Simplicial models
			and topological inference in biological systems},\ }in\ \href@noop {} {\emph
		{\bibinfo {booktitle} {Discrete and topological models in molecular
				biology}}}\ (\bibinfo  {publisher} {Springer},\ \bibinfo {year} {2014})\ pp.\
	\bibinfo {pages} {109--141}\BibitemShut {NoStop}%
	\bibitem [{\citenamefont {De~Silva}\ and\ \citenamefont
		{Ghrist}(2007)}]{de2007coverage}%
	\BibitemOpen
	\bibfield  {author} {\bibinfo {author} {\bibfnamefont {V.}~\bibnamefont
			{De~Silva}}\ and\ \bibinfo {author} {\bibfnamefont {R.}~\bibnamefont
			{Ghrist}},\ }\bibfield  {title} {\bibinfo {title} {Coverage in sensor
			networks via persistent homology},\ }\href@noop {} {\bibfield  {journal}
		{\bibinfo  {journal} {Algebraic \& Geometric Topology}\ }\textbf {\bibinfo
			{volume} {7}},\ \bibinfo {pages} {339} (\bibinfo {year} {2007})}\BibitemShut
	{NoStop}%
	\bibitem [{\citenamefont {Barbarossa}\ and\ \citenamefont
		{Tsitsvero}(2016)}]{barbarossa2016introduction}%
	\BibitemOpen
	\bibfield  {author} {\bibinfo {author} {\bibfnamefont {S.}~\bibnamefont
			{Barbarossa}}\ and\ \bibinfo {author} {\bibfnamefont {M.}~\bibnamefont
			{Tsitsvero}},\ }\bibfield  {title} {\bibinfo {title} {An introduction to
			hypergraph signal processing},\ }in\ \href@noop {} {\emph {\bibinfo
			{booktitle} {2016 IEEE International Conference on Acoustics, Speech and
				Signal Processing (ICASSP)}}}\ (\bibinfo {organization} {IEEE},\ \bibinfo
	{year} {2016})\ pp.\ \bibinfo {pages} {6425--6429}\BibitemShut {NoStop}%
	\bibitem [{\citenamefont {Pokorny}\ \emph {et~al.}(2016)\citenamefont
		{Pokorny}, \citenamefont {Hawasly},\ and\ \citenamefont
		{Ramamoorthy}}]{pokorny2016topological}%
	\BibitemOpen
	\bibfield  {author} {\bibinfo {author} {\bibfnamefont {F.~T.}\ \bibnamefont
			{Pokorny}}, \bibinfo {author} {\bibfnamefont {M.}~\bibnamefont {Hawasly}},\
		and\ \bibinfo {author} {\bibfnamefont {S.}~\bibnamefont {Ramamoorthy}},\
	}\bibfield  {title} {\bibinfo {title} {Topological trajectory classification
			with filtrations of simplicial complexes and persistent homology},\
	}\href@noop {} {\bibfield  {journal} {\bibinfo  {journal} {The International
				Journal of Robotics Research}\ }\textbf {\bibinfo {volume} {35}},\ \bibinfo
		{pages} {204} (\bibinfo {year} {2016})}\BibitemShut {NoStop}%
	\bibitem [{\citenamefont {Battiston}\ \emph {et~al.}(2020)\citenamefont
		{Battiston}, \citenamefont {Cencetti}, \citenamefont {Iacopini},
		\citenamefont {Latora}, \citenamefont {Lucas}, \citenamefont {Patania},
		\citenamefont {Young},\ and\ \citenamefont {Petri}}]{battiston2020networks}%
	\BibitemOpen
	\bibfield  {author} {\bibinfo {author} {\bibfnamefont {F.}~\bibnamefont
			{Battiston}}, \bibinfo {author} {\bibfnamefont {G.}~\bibnamefont {Cencetti}},
		\bibinfo {author} {\bibfnamefont {I.}~\bibnamefont {Iacopini}}, \bibinfo
		{author} {\bibfnamefont {V.}~\bibnamefont {Latora}}, \bibinfo {author}
		{\bibfnamefont {M.}~\bibnamefont {Lucas}}, \bibinfo {author} {\bibfnamefont
			{A.}~\bibnamefont {Patania}}, \bibinfo {author} {\bibfnamefont {J.-G.}\
			\bibnamefont {Young}},\ and\ \bibinfo {author} {\bibfnamefont
			{G.}~\bibnamefont {Petri}},\ }\bibfield  {title} {\bibinfo {title} {Networks
			beyond pairwise interactions: structure and dynamics},\ }\href@noop {}
	{\bibfield  {journal} {\bibinfo  {journal} {Physics Reports}\ }\textbf
		{\bibinfo {volume} {874}},\ \bibinfo {pages} {1} (\bibinfo {year}
		{2020})}\BibitemShut {NoStop}%
	\bibitem [{\citenamefont {Boguna}\ \emph {et~al.}(2021)\citenamefont {Boguna},
		\citenamefont {Bonamassa}, \citenamefont {De~Domenico}, \citenamefont
		{Havlin}, \citenamefont {Krioukov},\ and\ \citenamefont
		{Serrano}}]{boguna2021network}%
	\BibitemOpen
	\bibfield  {author} {\bibinfo {author} {\bibfnamefont {M.}~\bibnamefont
			{Boguna}}, \bibinfo {author} {\bibfnamefont {I.}~\bibnamefont {Bonamassa}},
		\bibinfo {author} {\bibfnamefont {M.}~\bibnamefont {De~Domenico}}, \bibinfo
		{author} {\bibfnamefont {S.}~\bibnamefont {Havlin}}, \bibinfo {author}
		{\bibfnamefont {D.}~\bibnamefont {Krioukov}},\ and\ \bibinfo {author}
		{\bibfnamefont {M.~{\'A}.}\ \bibnamefont {Serrano}},\ }\bibfield  {title}
	{\bibinfo {title} {Network geometry},\ }\href@noop {} {\bibfield  {journal}
		{\bibinfo  {journal} {Nature Reviews Physics}\ }\textbf {\bibinfo {volume}
			{3}},\ \bibinfo {pages} {114} (\bibinfo {year} {2021})}\BibitemShut {NoStop}%
	\bibitem [{\citenamefont {Mulder}\ and\ \citenamefont
		{Bianconi}(2018)}]{mulder2018network}%
	\BibitemOpen
	\bibfield  {author} {\bibinfo {author} {\bibfnamefont {D.}~\bibnamefont
			{Mulder}}\ and\ \bibinfo {author} {\bibfnamefont {G.}~\bibnamefont
			{Bianconi}},\ }\bibfield  {title} {\bibinfo {title} {Network geometry and
			complexity},\ }\href@noop {} {\bibfield  {journal} {\bibinfo  {journal}
			{Journal of Statistical Physics}\ }\textbf {\bibinfo {volume} {173}},\
		\bibinfo {pages} {783} (\bibinfo {year} {2018})}\BibitemShut {NoStop}%
	\bibitem [{\citenamefont {Anderson}\ and\ \citenamefont
		{May}(1992)}]{anderson1992infectious}%
	\BibitemOpen
	\bibfield  {author} {\bibinfo {author} {\bibfnamefont {R.~M.}\ \bibnamefont
			{Anderson}}\ and\ \bibinfo {author} {\bibfnamefont {R.~M.}\ \bibnamefont
			{May}},\ }\href@noop {} {\emph {\bibinfo {title} {Infectious diseases of
				humans: dynamics and control}}}\ (\bibinfo  {publisher} {Oxford University
		Press},\ \bibinfo {year} {1992})\BibitemShut {NoStop}%
	\bibitem [{\citenamefont {Bogun{\'a}}\ \emph {et~al.}(2013)\citenamefont
		{Bogun{\'a}}, \citenamefont {Castellano},\ and\ \citenamefont
		{Pastor-Satorras}}]{boguna2013nature}%
	\BibitemOpen
	\bibfield  {author} {\bibinfo {author} {\bibfnamefont {M.}~\bibnamefont
			{Bogun{\'a}}}, \bibinfo {author} {\bibfnamefont {C.}~\bibnamefont
			{Castellano}},\ and\ \bibinfo {author} {\bibfnamefont {R.}~\bibnamefont
			{Pastor-Satorras}},\ }\bibfield  {title} {\bibinfo {title} {Nature of the
			epidemic threshold for the susceptible-infected-susceptible dynamics in
			networks},\ }\href@noop {} {\bibfield  {journal} {\bibinfo  {journal}
			{Physical Review Letters}\ }\textbf {\bibinfo {volume} {111}},\ \bibinfo
		{pages} {068701} (\bibinfo {year} {2013})}\BibitemShut {NoStop}%
	\bibitem [{\citenamefont {Ferreira}\ \emph {et~al.}(2012)\citenamefont
		{Ferreira}, \citenamefont {Castellano},\ and\ \citenamefont
		{Pastor-Satorras}}]{ferreira2012epidemic}%
	\BibitemOpen
	\bibfield  {author} {\bibinfo {author} {\bibfnamefont {S.~C.}\ \bibnamefont
			{Ferreira}}, \bibinfo {author} {\bibfnamefont {C.}~\bibnamefont
			{Castellano}},\ and\ \bibinfo {author} {\bibfnamefont {R.}~\bibnamefont
			{Pastor-Satorras}},\ }\bibfield  {title} {\bibinfo {title} {Epidemic
			thresholds of the susceptible-infected-susceptible model on networks: A
			comparison of numerical and theoretical results},\ }\href@noop {} {\bibfield
		{journal} {\bibinfo  {journal} {Physical Review E}\ }\textbf {\bibinfo
			{volume} {86}},\ \bibinfo {pages} {041125} (\bibinfo {year}
		{2012})}\BibitemShut {NoStop}%
	\bibitem [{\citenamefont {Iacopini}\ \emph {et~al.}(2019)\citenamefont
		{Iacopini}, \citenamefont {Petri}, \citenamefont {Barrat},\ and\
		\citenamefont {Latora}}]{iacopini2019simplicial}%
	\BibitemOpen
	\bibfield  {author} {\bibinfo {author} {\bibfnamefont {I.}~\bibnamefont
			{Iacopini}}, \bibinfo {author} {\bibfnamefont {G.}~\bibnamefont {Petri}},
		\bibinfo {author} {\bibfnamefont {A.}~\bibnamefont {Barrat}},\ and\ \bibinfo
		{author} {\bibfnamefont {V.}~\bibnamefont {Latora}},\ }\bibfield  {title}
	{\bibinfo {title} {Simplicial models of social contagion},\ }\href@noop {}
	{\bibfield  {journal} {\bibinfo  {journal} {Nature Communications}\ }\textbf
		{\bibinfo {volume} {10}},\ \bibinfo {pages} {1} (\bibinfo {year}
		{2019})}\BibitemShut {NoStop}%
	\bibitem [{\citenamefont {Matamalas}\ \emph {et~al.}(2020)\citenamefont
		{Matamalas}, \citenamefont {G{\'o}mez},\ and\ \citenamefont
		{Arenas}}]{matamalas2020abrupt}%
	\BibitemOpen
	\bibfield  {author} {\bibinfo {author} {\bibfnamefont {J.~T.}\ \bibnamefont
			{Matamalas}}, \bibinfo {author} {\bibfnamefont {S.}~\bibnamefont
			{G{\'o}mez}},\ and\ \bibinfo {author} {\bibfnamefont {A.}~\bibnamefont
			{Arenas}},\ }\bibfield  {title} {\bibinfo {title} {Abrupt phase transition of
			epidemic spreading in simplicial complexes},\ }\href@noop {} {\bibfield
		{journal} {\bibinfo  {journal} {Physical Review Research}\ }\textbf {\bibinfo
			{volume} {2}},\ \bibinfo {pages} {012049} (\bibinfo {year}
		{2020})}\BibitemShut {NoStop}%
	\bibitem [{\citenamefont {Kovalenko}\ \emph {et~al.}(2021)\citenamefont
		{Kovalenko}, \citenamefont {Sendi{\~n}a-Nadal}, \citenamefont {Khalil},
		\citenamefont {Dainiak}, \citenamefont {Musatov}, \citenamefont
		{Raigorodskii}, \citenamefont {Alfaro-Bittner}, \citenamefont {Barzel},\ and\
		\citenamefont {Boccaletti}}]{growing}%
	\BibitemOpen
	\bibfield  {author} {\bibinfo {author} {\bibfnamefont {K.}~\bibnamefont
			{Kovalenko}}, \bibinfo {author} {\bibfnamefont {I.}~\bibnamefont
			{Sendi{\~n}a-Nadal}}, \bibinfo {author} {\bibfnamefont {N.}~\bibnamefont
			{Khalil}}, \bibinfo {author} {\bibfnamefont {A.}~\bibnamefont {Dainiak}},
		\bibinfo {author} {\bibfnamefont {D.}~\bibnamefont {Musatov}}, \bibinfo
		{author} {\bibfnamefont {A.~M.}\ \bibnamefont {Raigorodskii}}, \bibinfo
		{author} {\bibfnamefont {K.}~\bibnamefont {Alfaro-Bittner}}, \bibinfo
		{author} {\bibfnamefont {B.}~\bibnamefont {Barzel}},\ and\ \bibinfo {author}
		{\bibfnamefont {S.}~\bibnamefont {Boccaletti}},\ }\bibfield  {title}
	{\bibinfo {title} {Growing scale-free simplices},\ }\href@noop {} {\bibfield
		{journal} {\bibinfo  {journal} {Communication Physics}\ }\textbf {\bibinfo
			{volume} {4}},\ \bibinfo {pages} {43} (\bibinfo {year} {2021})}\BibitemShut
	{NoStop}%
\end{thebibliography}
\providecommand{\noopsort}[1]{}\providecommand{\singleletter}[1]{#1}%

\end{document}